\documentclass[iop]{emulateapj}
\usepackage[pass,letterpaper]{geometry}
\usepackage{hyperref}

\journalinfo{Astrophysical Journal in press}
\submitted{Received 2014 August 25; accepted 2014 September 15}%; published 2014 YYYYYYY XX}

\shorttitle{The Best and Brightest Metal-Poor Stars}
\shortauthors{Schlaufman \& Casey}

\begin{document}

%\title{The Best and Brightest Metal-Poor Stars\footnote{This paper
%includes data gathered with the 6.5 meter Magellan Telescopes located
%at Las Campanas Observatory, Chile.}}
\title{THE BEST AND BRIGHTEST METAL-POOR STARS\footnote{This paper
includes data gathered with the 6.5 meter Magellan Telescopes located
at Las Campanas Observatory, Chile.}}
\title{THE BEST AND BRIGHTEST METAL-POOR STARS\altaffilmark{1}}

\author{Kevin C.\ Schlaufman\altaffilmark{2,3,4} \& Andrew R.\ Casey\altaffilmark{2,4,5,6}}
\email{kschlauf@mit.edu, arc@ast.cam.ac.uk}

\altaffiltext{1}{This paper includes data gathered with the 6.5 meter
Magellan Telescopes located at Las Campanas Observatory, Chile.}
\altaffiltext{2}{Kavli Institute for Astrophysics and Space Research,
Massachusetts Institute of Technology, Cambridge, MA 02139, USA}
\altaffiltext{3}{Kavli Fellow.}
\altaffiltext{4}{Visiting Astronomer, Kitt Peak National Observatory,
National Optical Astronomy Observatory, which is operated by the
Association of Universities for Research in Astronomy (AURA) under
cooperative agreement with the National Science Foundation.}
\altaffiltext{5}{Research School of Astronomy and Astrophysics, Australian
National University, Canberra, ACT 2611, Australia}
\altaffiltext{6}{Institute of Astronomy, University of Cambridge,
Madingley Road, Cambridge CB3 0HA, UK}

\begin{abstract}

The chemical abundances of large samples of extremely metal-poor
(EMP) stars can be used to investigate metal-free stellar populations,
supernovae, and nucleosynthesis as well as the formation and galactic
chemical evolution of the Milky Way and its progenitor halos.  However,
current progress on the study of EMP stars is being limited by their faint
apparent magnitudes.  The acquisition of high signal-to-noise spectra
for faint EMP stars requires a major telescope time commitment, making
the construction of large samples of EMP star abundances prohibitively
expensive.  We have developed a new, efficient selection that uses only
public, all-sky APASS optical, 2MASS near-infrared, and WISE mid-infrared
photometry to identify bright metal-poor star candidates through their
lack of molecular absorption near 4.6 microns.  We have used our selection
to identify 11,916 metal-poor star candidates with $V<14$, increasing
the number of publicly-available candidates by more than a factor of
five in this magnitude range.  Their bright apparent magnitudes have
greatly eased high-resolution follow-up observations that have identified
seven previously unknown stars with $\mathrm{[Fe/H]}\lesssim-3.0$.
Our follow-up campaign has revealed that $3.8^{+1.3}_{-1.1}$\% of our
candidates have $\mathrm{[Fe/H]}\lesssim-3.0$ and $32.5^{+3.0}_{-2.9}$\%
have $-3.0\lesssim\mathrm{[Fe/H]}\lesssim-2.0$.  The bulge is the
most likely location of any existing Galactic Population III stars,
and an infrared-only variant of our selection is well suited to the
identification of metal-poor stars in the bulge.  Indeed, two of our
confirmed metal-poor stars with [Fe/H] $\lesssim-2.7$ are within about
2 kpc of the Galactic Center.  They are among the most metal-poor stars
known in the bulge.

\end{abstract}


\section{Introduction}

The chemical abundances of extremely metal-poor (EMP) stars
uniquely illuminate the chemical state of star-forming regions in
the progenitor halos of the Milky Way.  The relatively low-mass EMP
stars still observable today must have formed in star-forming regions
seeded with metals by more massive and more primitive stars, perhaps
even by metal-free Population III stars.  As a result, the chemical
abundances of large samples of EMP stars collectively constrain the high
redshift initial mass function, the nucleosynthetic yields and explosive
deaths of Population III stars, the production of lithium from Big Bang
nucleosynthesis, as well as the galactic chemical evolution of the Milky
Way \citep[e.g.,][]{yos06,heg10,rya99,cay04,bee05}.

Classical objective prism surveys for metal-poor stars like the HK Survey
of \citet{bee85,bee92} identified about 1,000 candidates with $V \lesssim
14$ based on weak Ca II H \& K absorption lines.  More recently, the
objective prism-based Hamburg/ESO Survey (HES) has identified more than
20,000 candidates using Ca II H \& K, but typically with $V \gtrsim 16$
\citep{wis96,rei97,chr08}.  Moderate-resolution follow-up observations
of candidates from both the HK Survey and HES are still in progress.
Taking advantage of the fact that a significant fraction of EMP stars
are carbon enhanced, these carbon-enhanced EMP stars (CEMP) stars can
also be identified by prominent carbon features in objective-prism data
\citep[e.g.,][]{pla10,pla11}.

The importance of identifying apparently bright metal-poor stars led
\citet{fre06b} to perform an independent search of the HES plates for
candidates with $9 \lesssim B \lesssim 14$.  That search and subsequent
moderate-resolution spectroscopy identified about 150 stars with $-3.0
\lesssim \mathrm{[Fe/H]} \lesssim -2.0$ and approximately 25 stars with
$-4.0 \lesssim \mathrm{[Fe/H]} \lesssim -3.0$.  The search also led to
the discovery of HE 1327-2326, at the time the most iron-poor star known
\citep{fre05,aok06,fre06a}.  High-resolution follow-up observations of
this sample are ongoing \citep[e.g.,][]{hol11}.

The Sloan Digital Sky Survey \citep[SDSS;][]{yor00} and the
subsequent Sloan Extension for Galactic Understanding and Exploration
\citep[SEGUE;][]{yan09} have also led to the discovery of many
EMP stars.  Searches based on the weakness of Ca II H \& K in SDSS
$R \sim 1,\!800$ spectra have confirmed tens of EMP stars with $g
\gtrsim 16$, including the star with the lowest-known total metallicity
\citep[e.g.,][]{caf11a,caf11b,caf12,caf13,bon12}.  At the same time,
metallicities directly determined from SDSS spectroscopy have led
to the confirmation of about 100 EMP stars and the identification of
about 1,000 very likely EMP stars \citep[e.g.,][]{all06,aok13,nor13a}.
As with objective prism data, prominent carbon features observed in
SDSS spectroscopy have led to the confirmation of 24 EMP stars and the
identification of about 50 very likely EMP stars with significant carbon
enhancements \citep[e.g.,][]{aok08,lee13,spi13}.  The RAdial Velocity
Experiment \citep[RAVE;][]{ste06} was an untargeted spectroscopic survey
of bright stars over 20,000 square degrees of the southern sky that has
also identified more than 1,000 stars with $\mathrm{[Fe/H]} \lesssim
-2.0$ \citep[e.g.,][]{ful10,kor13}

The 1.3m SkyMapper Telescope at Siding Spring Observatory is poised to
produce a flood of new EMP star candidates.  The combination of its
$ugriz$ filter set with a $v$ intermediate-band filter covering the
spectrum from 367 nm to 398 nm is optimized to identify metal-poor stars
\citep{kel07}.  The SkyMapper-EMP star program is expected to identify
a further $\sim\!\!5,\!000$ candidates with $9 < g < 15$ and another
$\sim\!\!26,\!000$ candidates with $15 < g < 17$.  Already, SkyMapper
has discovered the most iron-poor star known: SMSS J031300.36-670839.3
\citep{kel14}.  The chemical abundance pattern of SMSS J031300.36-670839.3
suggests that it likely formed from material enriched by a only a single
supernova, indicating that it is the closest link to Population III
stars yet found.

Even with tens of thousands of metal-poor star candidates known, their
faint apparent magnitudes leave their enormous scientific potential
tantalizingly close yet just out of reach.  High-resolution spectroscopy
(e.g., $R \gtrsim 25,\!000$) with high signal-to-noise (e.g., S/N $\sim
100$/pixel at 400 nm) is critical to fully understand and exploit the
chemical abundances of EMP stars.  Yet to achieve such spectroscopy
takes a prohibitively large amount of 6--10 m class telescope time.
For example, the acquisition of a $R \sim 25,\!000$ spectrum with
S/N $\sim 100$/pixel at 400 nm for an EMP star with $V \approx 16$
takes about four hours using the MIKE spectrograph on the Magellan
Clay Telescope.  The same resolution and S/N could be achieved for
a $V \approx 13$ EMP star in only 15 minutes.  The lack of apparently
bright metal-poor stars has restricted the most comprehensive metal-poor
star abundance analyses yet published to only a few hundred objects
\citep{nor13a,nor13b,yon13a,yon13b,roe14}.

To help identify large samples of apparently bright EMP stars, we
have developed a new, efficient metal-poor star candidate selection
that uses only public, all-sky APASS optical, 2MASS near-infrared,
and WISE mid-infrared photometry.  Metal-rich stars are bright in the
2MASS $J$ and WISE $W1$ bands, but faint in the WISE $W2$ band due to
molecular absorption.  Metal-poor stars are bright in all three bands.
We specify our sample selection in Section 2.  We detail our observational
follow up, stellar parameter derivation, and carbon abundance analysis in
Section 3.  We discuss our results and their implications in Section 4,
and we summarize our findings in Section 5.

\section{Sample Selection}

As input to our sample selection, we use the APASS DR6 Catalog, the
2MASS All-Sky Point Source Catalog, and the AllWISE Source Catalog
\citep{hen12,skr06,wri10,mai11}.  We describe the data quality cuts we
apply in the Appendix.

\begin{figure*}
%\epsscale{1.0} % aastex
\plotone{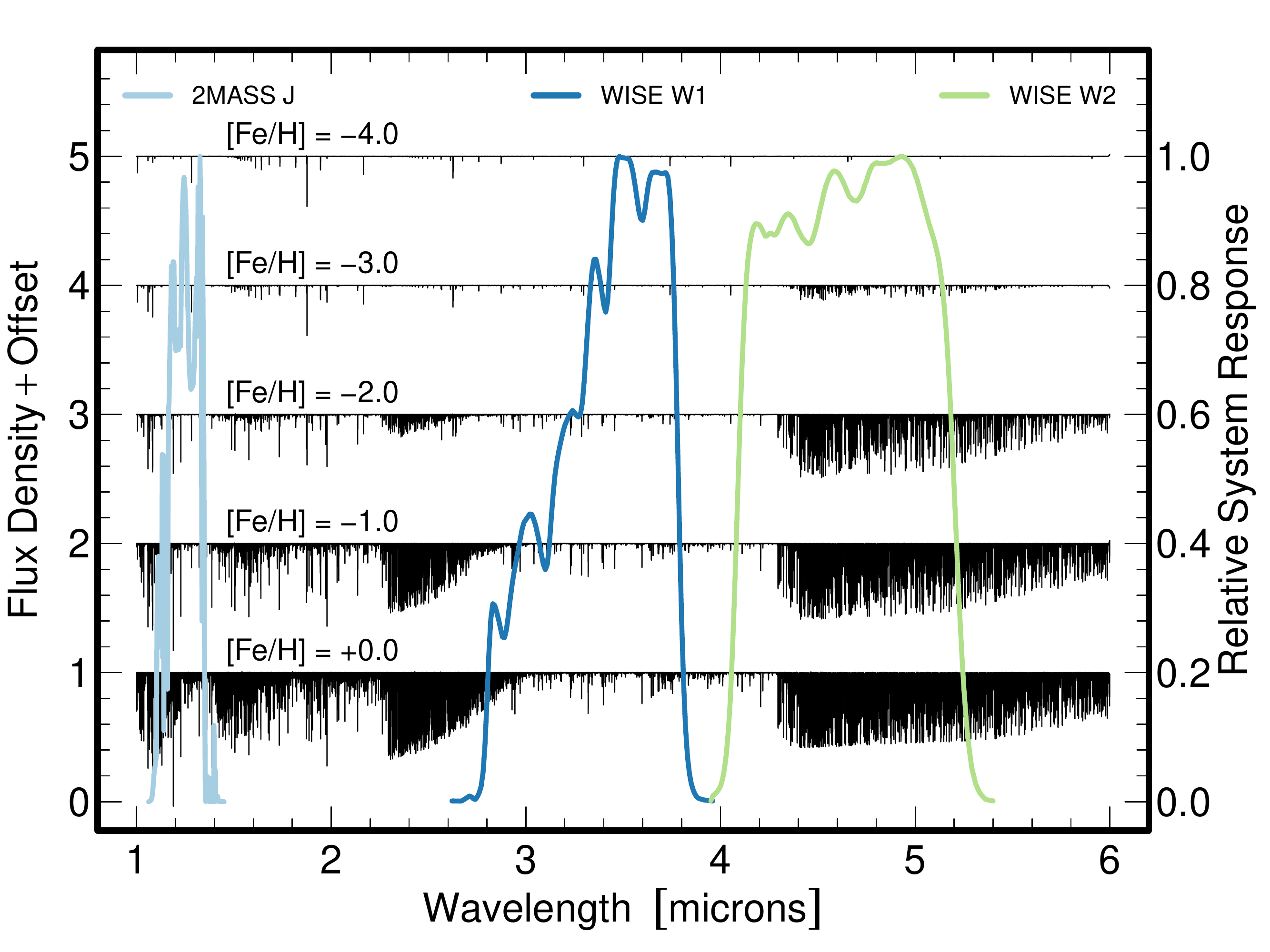}
\caption{\citet{bro05} theoretical spectra for stellar atmospheres with
the metallicity given in the plot, assuming [$\alpha$/Fe] = +0.4 for
[Fe/H] $\leq -1$.  We assume $T_{\mathrm{eff}} = 4800$ K -- the median
of our sample -- and $\log{g} = 1.5$, though the features are insensitive
to $\log{g}$.  The light blue curve is the relative system response curve
(RSR) for the 2MASS $J$ band ($1.2~\mu$), the dark blue curve is the RSR
for the WISE $W1$ band ($3.4~\mu$), and the green curve is the RSR for
the WISE $W2$ band ($4.6~\mu$).  Strong molecular absorption is present
in $W2$ down to [Fe/H] $= -2$, implying that colors involving $W2$
can help select metal-poor stars from photometry alone.\label{fig01}}
\end{figure*}

We show the physical basis for our selection in Figure~\ref{fig01}.
Strong molecular absorption is present in the atmospheres of stars with
effective temperature $T_{\mathrm{eff}}$ in the range $4500~\mathrm{K}
\lesssim T_{\mathrm{eff}} \lesssim 5500$ K for all surface gravities
$\log{g}$, strongly affecting fluxes in the WISE $W2$ band at 4.6 microns.
While this strong absorption persists down to [Fe/H] $\sim -2.0$
in cool stars, it disappears at higher metallicity for hotter stars.
As a result, we use 2MASS $J-H$ color as a reddening-insensitive proxy
for effective temperature and select stars with $0.45 \leq J-H \leq 0.6$.
We also eliminate nearly all solar neighborhood interlopers by selecting
objects with $W3 > 8$.

To make the qualitative observation that $W2$ flux can be used
to identify metal-poor stars more quantitative, we download all
stars from SIMBAD with published stellar parameters in the range
$3000~\mathrm{K} < T_{\mathrm{eff}} < 8000$ K and cross-match the result
with the 2MASS All-Sky Point Source and AllWISE Source catalogs using
\texttt{TOPCAT}\footnote{\url{http://www.star.bristol.ac.uk/~mbt/topcat/}}
\citep{tay05}.  WISE photometry is in Vega magnitudes, so a star with a
SED similar to Vega will have all colors composed of WISE filters equal
to zero.  Vega is a hot star with no molecular absorption in $W2$,
so metal-poor stars with no molecular absorption in $W2$ should have
similar SEDs to Vega and have $W1 - W2 \approx 0$.  More metal-rich stars
will have slightly bluer $W1-W2$ colors, so we require $-0.04 \leq W1-W2
\leq 0.04$.  Empirically, we find that selecting stars with $J-W2 > 0.5$
has a small positive impact on the yield of metal-poor stars by removing
a significant number of solar-metallicity contaminants.

We used this pure color selection for our initial target catalog, then
used the metal-poor stars identified on each observing run to optimize
the selection in three more ways.  First, we used the metal-poor stars
we identified to train a logistic regression model\footnote{See Section
3.1 of \citet{sch14} for a detailed discussion of logistic regression.}
to predict the probability that a star that passes our color cuts
is metal-poor based on its $J-H$, $J-K$, $J-W1$, and $J-W2$ colors.
Second, we noted that the metal-poor stars we discovered had red $J-W2$
colors at constant $B-V$.  Third, we noted that none of our metal-poor
stars had very red $B-V$ colors, so we require $B-V < 1.2$. In summary,
our final list of selection criteria is:

\begin{enumerate}
\item
$0.45 \leq J-H \leq 0.6$
\item
$W3 > 8$
\item
$-0.04 \leq W1-W2 \leq 0.04$
\item
$J-W2 > 0.5$
\item
$e^{z}/(1+e^{z}) > 0.13$, where $z = -2.534642 + 16.241145
\left(J-H\right) - 9.271496 \left(J-K\right) - 40.009841 \left(J-W1\right)
+ 38.514156 \left(J-W2\right)$
\item
$J-W2 > 0.5 \left[\left(B-V\right)-0.8\right]+0.6$
\item
$B-V < 1.2$
\end{enumerate}

We estimate our completeness by calculating the fraction of stars in the
compilation of \citet{yon13a} that we would recover with our selection.
Of the 29 stars from \citet{yon13a} that pass our WISE data quality checks
and are in the range $0.45 \leq J-H \leq 0.6$, criteria (1)--(5) identify
18 (62\%) as metal-poor candidates with infrared photometry alone.
Of the 16 stars that pass criteria (1)--(5) and that have APASS data,
14 (88\%) are identified as metal poor by our full selection.  We also
confirm that our selection correctly identifies SMSS J031300.36-670839.3
as a metal-poor candidate.  After applying criteria (1)--(5), we are left
with 22,721 metal-poor star candidates with $V < 14$.  Applying our full
selection leaves us with 11,916 metal-poor star candidates with $V < 14$.

We subsequently found that most of our false positives are stars
with emission in the cores of the Ca II H \& K lines as well as broad
absorption lines demonstrating significant projected rotation velocities.
These objects are likely young stars, as stars in the range of $J-H$ we
explore spin-down to slower rotation velocities within $\approx\!\!650$
Myr \citep{irw09}.  Since our selection is based on an excess of $W2$
flux relative to that expected for solar-metallicity stars, these stars
are selected because they are orbited by hot debris disks.  The occurrence
of debris declines sharply with age for stars in this range of color,
so hot debris is another signal of youth \citep[e.g.,][]{pla05,rhe07}.
For all of these reasons, our selection would also be very useful to
select young stars in the field.  The addition of soft X-ray data to
our selection from the upcoming all-sky X-ray survey by the eROSITA
satellite will likely enable a clean separation of metal-poor stars and
young stars \citep{mer12}.

\section{Observational Follow-Up and Determination of Stellar Parameters}

We followed up our metal-poor star candidates with the Mayall
4m/Echelle\footnote{Program 2013A-0186.}, Gemini South/GMOS-S
\citep{hoo04}\footnote{Programs GS-2014A-Q-8 and GS-2014A-Q-74.}, and
Magellan/MIKE \citep{ber03} telescopes and spectrographs.  We observed 98
stars with the Mayall 4m/Echelle on 25--27 June 2013, using the 58.5-63
grating, cross-disperser 226-2, 1.0\arcsec~slit, order 2, and the blue
camera, corrector, and collimator.  While we found no EMP stars on our
Mayall 4m/Echelle run, we did use the data to significantly improve our
selection.  We do not consider it further.  We observed 90 stars with
Gemini South/GMOS-S in service mode from March to July 2014.  We used
the long-slit mode with the B1200 grating, 0.5\arcsec~slit, and central
wavelength 450 nm, yielding resolution $R \approx 3,\!700$ spectra between
380 nm and 520 nm.  We observed 416 stars with Magellan/MIKE on 21--23
June and 8--10 July 2014.  We used the 0.7\arcsec~slit and the standard
blue and red grating azimuths, yielding resolution $R \approx 25,\!000$
spectra between 335 nm and 950 nm.

\subsection{Gemini South/GMOS-S Data}

We derive stellar parameters for the stars we observed with Gemini
South/GMOS-S by comparing the observed spectra with model spectra in
a Markov Chain Monte Carlo (MCMC) framework.  Our code creates model
spectra by interpolating from the high-resolution AMBRE synthetic
library \citep{lav12} for each sampling of $T_{\mathrm{eff}}$, $\log{g}$,
[Fe/H], and mean $\alpha$-element enhancement [$\alpha$/Fe].  Additional
phenomena that transform the observed spectra were included as free
parameters: radial velocity, continuum coefficients, broadening, and
outlier pixels.  There are 11 total model parameters: $T_{\mathrm{eff}}$,
$\log{g}$, [Fe/H], [$\alpha$/Fe], redshift $z$, continuum coefficients
$\{c_i\}_{i=1}^{3}$, logarithm of fractionally underestimated variance
$\log{f}$, outlier fraction $P_o$, and additive outlier variance $V_o$.

We numerically optimized the model parameters
before performing the MCMC analysis using
\texttt{emcee}\footnote{\url{http://dan.iel.fm/emcee/current/}}
\citep{for13}.  We took the starting point for 200 walkers as a
multi-dimensional ball around the optimized point.  We allowed for $10^6$
total steps for the burn-in phase, and we used a further $2\times10^5$
steps to sample the posterior.  The auto-correlation times, mean
acceptance fractions, and values of the model parameters with each step
suggest that the analyses are converged by at most $3\times10^5$ steps.
We give the mean of the posterior and credible intervals for stellar
parameters and radial velocity in Table~\ref{tbl-1}.

\subsection{Magellan/MIKE Data}

We derive stellar parameters for the stars we observed with Magellan/MIKE
by classical methods.  Individual echelle orders were normalized with
a spline function before stitching them together, forming a contiguous
spectrum from $335-950$ nm.  Our spectra typically have S/N $\approx
50$/pixel at 550 nm.  We determined radial velocities by cross correlation
with the rest-frame spectrum of a metal-poor giant, then Doppler shifted
the data to the rest frame.  We measured equivalent widths of Fe I and
II lines by fitting Gaussian profiles using the approach described by
\citet{cas14}.  All transition measurements were visually inspected and we
discarded poorly fit lines.  We sourced our atomic data from \citet{roe10}
and our molecular data (CH) from \citet{ple08}.  We used \citet{cas04}
$\alpha$-enhanced model atmospheres.

We estimated our stellar parameters through excitation and ionization
balance.  We found $T_{\mathrm{eff}}$ by enforcing a zero trend in Fe
I abundance with excitation potential.  We adjusted $\log{g}$ until the
mean Fe I and Fe II abundances matched within 0.005 dex.  Similarly, we
adjusted the model atmosphere until it matched the mean Fe I abundance.
Finally, we found microturbulence $\xi$ by ensuring a zero trend in
abundance and reduced equivalent width for all Fe I lines. These four
conditions were simultaneously met to derive stellar parameters for
the entire Magellan/MIKE sample, which we give in Table~\ref{tbl-2}.
Given the spectral resolution and S/N of our sample, the uncertainties
in stellar parameters for the Magellan/MIKE sample are conservatively
estimated to be of order 100 K in $T_{\rm eff}$, 0.2 in $\log{g}$, 0.1 in
[Fe/H], and 0.1 km s$^{-1}$ in $\xi$.  Some of our candidates proved to be
so metal rich that an equivalent width analysis was inappropriate.  We do
not report stellar parameters for these objects and express the reason
in the comment column of Table~\ref{tbl-2}.  We give the photometry for
our Magellan/MIKE sample in Table~\ref{tbl-3}.  We selected our targets
from our larger catalog because their equatorial coordinates made them
easy to observe at low air mass from Las Campanas in June and July.
We also focused on stars with $V \lesssim 13.5$.

\section{Discussion}

We observed 239 stars selected using all seven criteria on our full
APASS, 2MASS, and WISE dataset.  We designate these stars by ``v2"
in the selection column in Table~\ref{tbl-3}.  We identified four EMP
stars with [Fe/H] $\lesssim -3.0$ and 73 very metal-poor (VMP) stars with
$-3.0 \lesssim \mathrm{[Fe/H]} \lesssim -2.0$.  In our candidate list,
there were also five EMP stars and nine VMP stars that had already been
discovered and that we would have observed had they been unknown.  We list
these stars in Table~\ref{tbl-4}.  We therefore estimate the efficiency of
our selection as 9/253 for EMP stars and 82/253 for VMP stars, indicating
fractional Bayesian selection efficiencies\footnote{See Section 3.3 of
\citet{sch14} for a detailed discussion.} of $3.8^{+1.3}_{-1.1}$\% and
$32.5^{+3.0}_{-2.9}$\%.  We observed 173 stars selected using criteria
(1)--(5) on our pure infrared 2MASS and WISE dataset.  We designate
these stars by ``v1" in the selection column in Table~\ref{tbl-3}.
We identified 3 EMP stars and 35 VMP stars.  We therefore estimate
the efficiency of our selection as 3/173 for EMP stars and 35/173
for VMP stars, indicating fraction Bayesian selection efficiencies of
$2.1^{+1.3}_{-1.0}$\% and $20.5^{+3.2}_{-2.9}$\%.

\begin{figure*}
%\epsscale{0.8} % aastex
\plotone{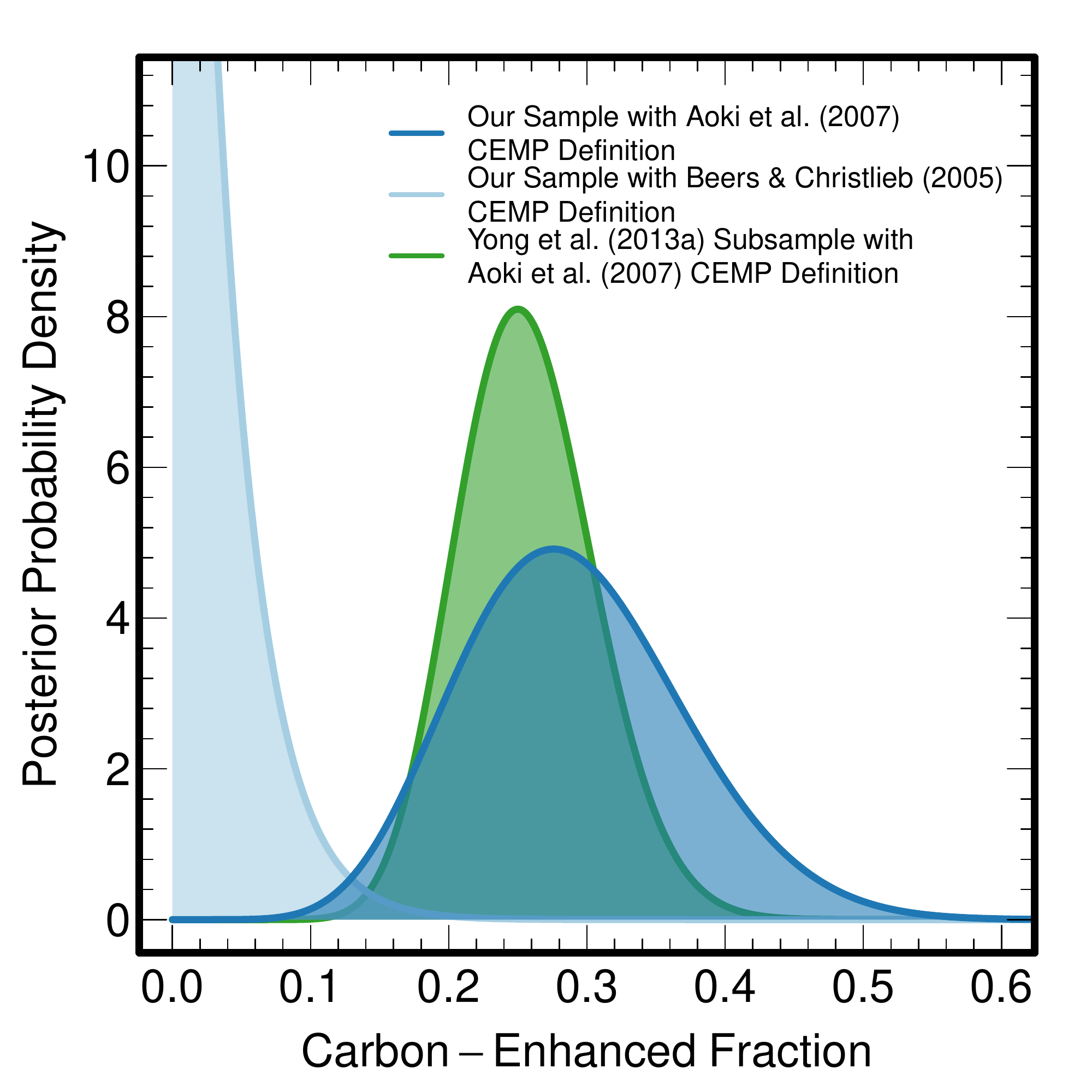}
\caption{Posterior probability density of our carbon-enhanced
fraction estimate.  The dramatic difference for our sample between the
\citet{bee05} and the \citet{aok07} definitions of carbon enhancement
demonstrates the importance of the \citet{aok07} definition's allowance
for stellar evolution.  The carbon-enhanced fraction in our sample of
29 stars with $-3.1 \leq \mathrm{[Fe/H]} \leq -2.5$ is fully consistent
with the carbon-enhanced fraction in a subsample of stars with $-3.1
\leq \mathrm{[Fe/H]} \leq -2.5$ from \citet{yon13a}.\label{fig02}}
\end{figure*}

We plot in Figure~\ref{fig02} the posterior probability distribution for
the fraction of carbon enhanced stars in our Magellan/MIKE sample of stars
with $\mathrm{[Fe/H]} \leq -2.5$.  We use both the original definition of
carbon enhancement $\mathrm{[C/Fe]} > +1.0$ given in \citet{bee05} and the
update given in \citet{aok07} accounting for stellar evolutionary effects

\begin{enumerate}
\item
$\mathrm{[C/Fe]}\geq+0.7$ for stars with $\log{L/L_{\odot}}\leq2.3$,
\item
$\mathrm{[C/Fe]}\geq+3.0-\log{L/L_{\odot}}$ for stars with
$\log{L/L_{\odot}}>2.3$.
\end{enumerate}

\noindent
Of our 29 stars with $\mathrm{[Fe/H]} \leq -2.5$, zero are carbon enhanced
by the \citet{bee05} definition, while eight are carbon enhanced by the
\citet{aok07} definition.  Carbon in the atmospheres of low surface
gravity stars is mixed into the stellar interior, where it can be
processed into nitrogen \citep[e.g.,][]{cha95}.  The expected enhancement
of nitrogen in the atmospheres of low surface gravity metal-poor stars has
been observationally confirmed \citep[e.g.,][]{gra00,spi05}.  The dramatic
difference in the carbon enhanced fraction between the two definitions
emphasizes the importance of using the \citet{aok07} definition to
investigate carbon enhancement in stars above the horizontal branch.

\begin{figure*}
%\epsscale{0.8} % aastex
\plotone{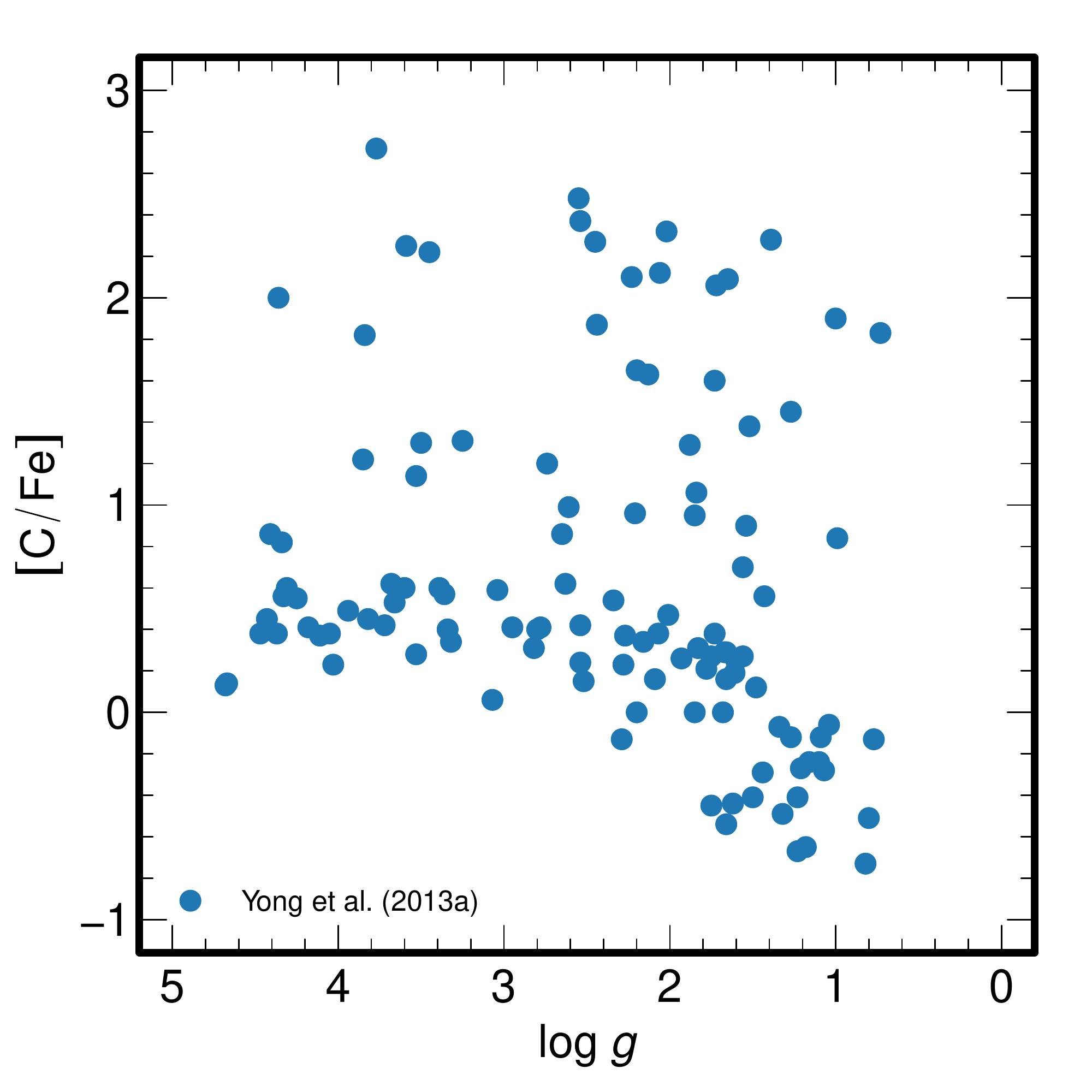}
\caption{[C/Fe] versus $\log{g}$ from \citet{yon13a}.  At $\log{g} \approx
1.5$ typical of our sample, $\mathrm{[C/Fe]} \lesssim 2.0$.\label{fig03}}
\end{figure*}

\begin{figure*}
%\epsscale{0.8} % aastex
\plotone{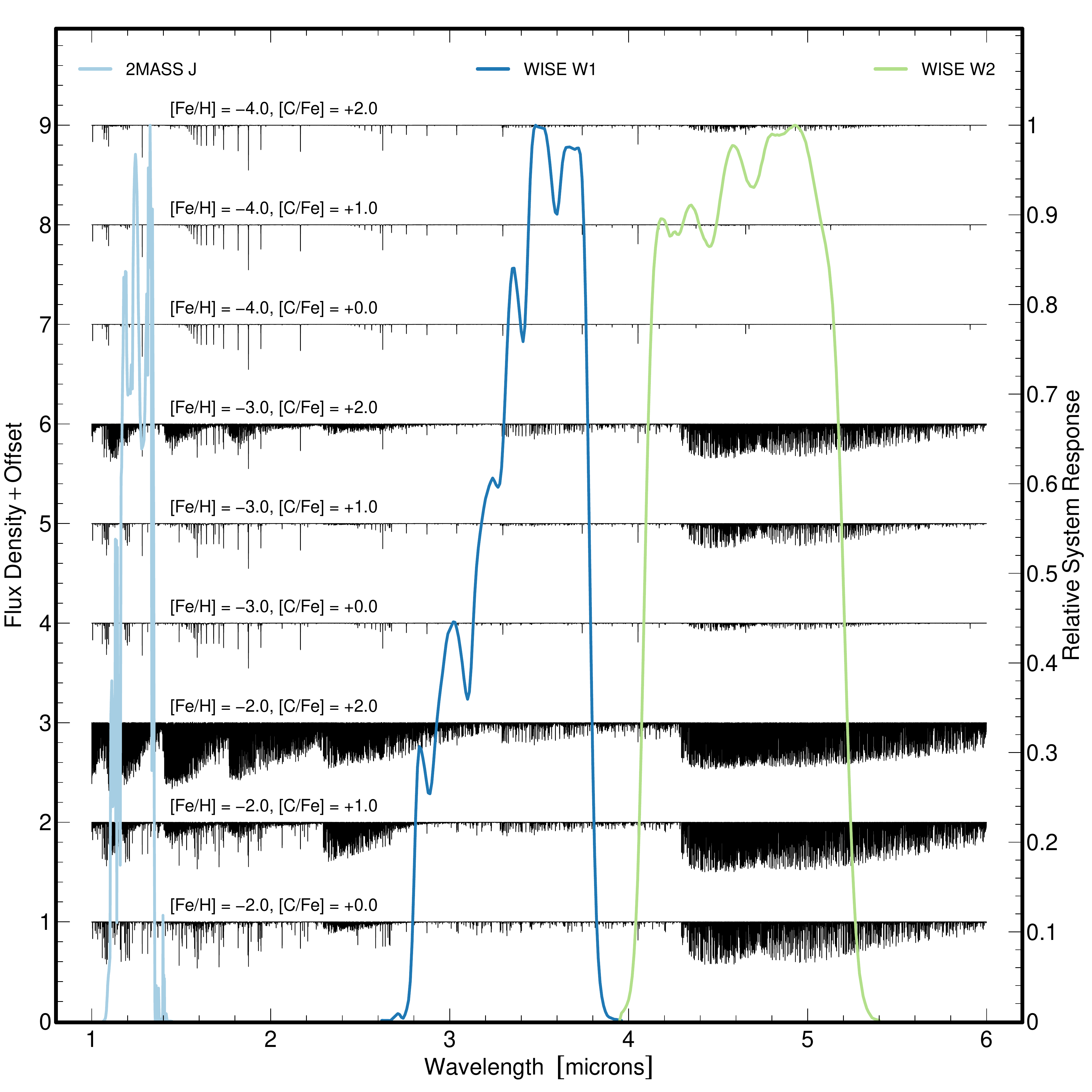}
\caption{Theoretical spectra for stellar atmospheres with the metallicity
and carbon enhancement given in the plot, assuming [$\alpha$/Fe] = +0.4
(T.~Masseron, private communication).  We assume $T_{\mathrm{eff}} =
4800$ K -- the median of our sample -- and $\log{g} = 1.5$, though the
features are insensitive to $\log{g}$.  The light blue curve is the RSR
for the 2MASS $J$ band ($1.2~\mu$), the dark blue curve is the RSR for
the WISE $W1$ band ($3.4~\mu$), and the green curve is the RSR for the
WISE $W2$ band ($4.6~\mu$).  Our selection correctly separates stars
with $\mathrm{[Fe/H]} \approx -2.0$ and $\mathrm{[C/Fe]} \approx +0.0$
from the field population based on the weakness of molecular absorption
in $W2$.  An EMP star with $\mathrm{[C/Fe]} \approx +2.0$ still has weaker
absorption in $W2$ than a star with $\mathrm{[Fe/H]} \approx -2.0$ and
$\mathrm{[C/Fe]} \approx +0.0$, so our selection is unbiased against
carbon-enhanced EMP stars.  It is possible that our selection could
miss stars with $\mathrm{[Fe/H]} \approx -2.0$ and $\mathrm{[C/Fe]}
\gtrsim +0.0$.\label{fig04}}
\end{figure*}

We plot in Figure~\ref{fig03} the distribution of metal-poor stars
from \citet{yon13a} with measured [C/Fe] in the $\log{g}$--[C/Fe]
plane.  Evolved metal-poor stars like those in our sample typically
have $\mathrm{[C/Fe]} \lesssim +2.0$.  We plot in Figure~\ref{fig04}
carbon-enhanced synthetic spectra from 1 to 6 microns.  Our search
efficiently selects stars with $\mathrm{[Fe/H]} \approx -2.0$ and
$\mathrm{[C/Fe]} \approx +0.0$.  Since even a very carbon-enhanced EMP
star with $\mathrm{[Fe/H]} \approx -3.0$ and $\mathrm{[C/Fe]} \approx
+2.0$ has weaker absorption in $W2$ than the aforementioned carbon-normal
very metal-poor star, our search technique itself is not biased against
carbon-enhanced EMP stars.  While targeting carbon enhanced stars
can increase the yield from EMP star searches, it can complicate the
interpretation of the carbon-enhanced fraction.  Conversely, further
exploration using our selection will be able to provide an unbiased
estimate of the carbon-enhanced fraction as a function of metallicity.

\begin{figure*}
%\epsscale{0.8}	% aastex
\plotone{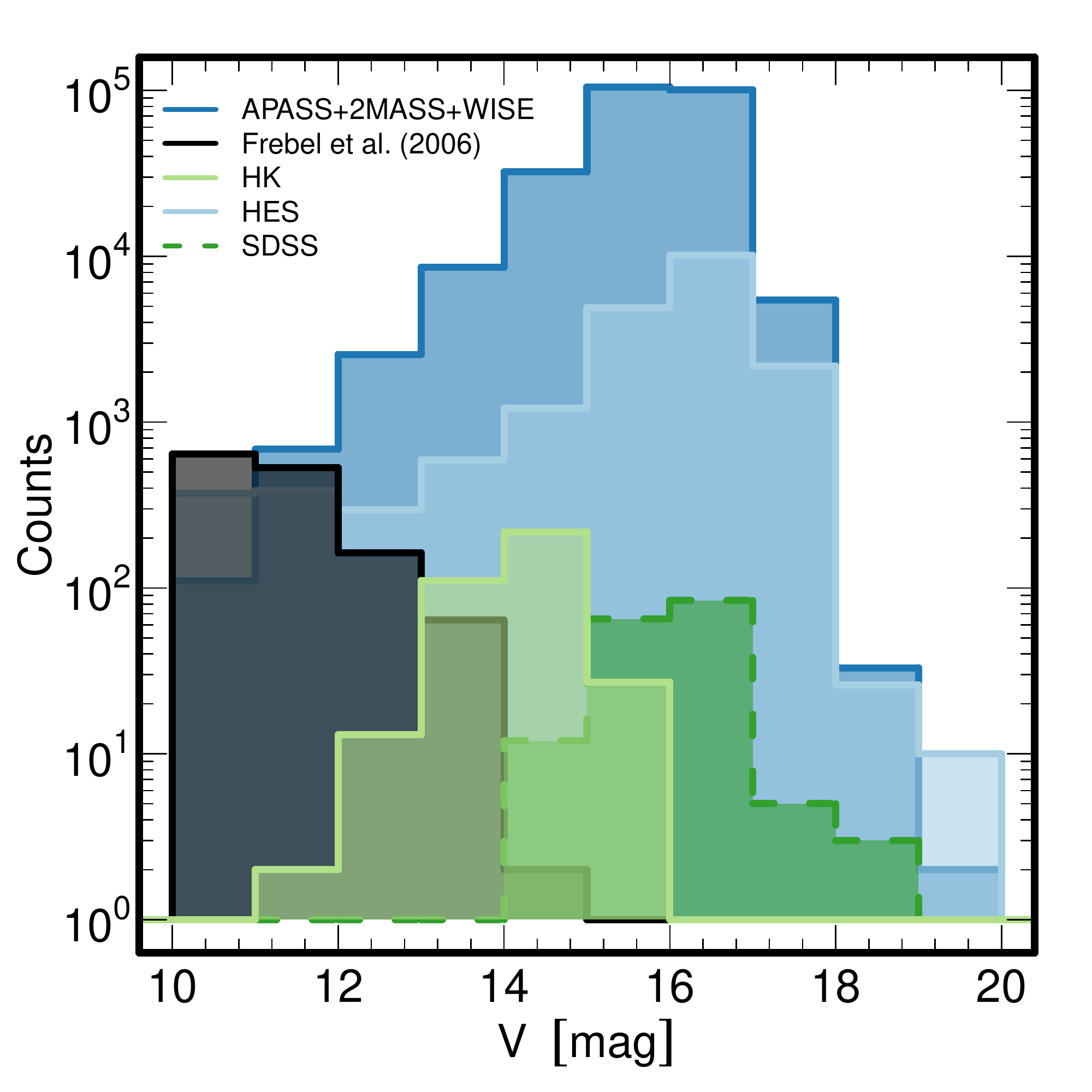}
\caption{$V$-band magnitude distributions of candidate metal-poor
stars.  We plot the distributions from our selection (dark blue),
the \citet{fre06b} survey (black), the HK Survey \citep[light
green;][]{bee92}, and the Hamburg/ESO Survey \citep[light
blue;][]{chr08}.  For EMP searches based on the SDSS, we plot the
apparent magnitude distribution of confirmed EMP stars \citep[dark
green;][]{aok08,aok13,bon12,caf11b,caf12,caf13}.  That distribution
is biased to bright apparent magnitudes compared to the candidate
distribution, so the SDSS distribution should be understood as a bright
limit to the SDSS metal-poor candidate distribution.  For the SDSS sample,
we have calculated approximate $V$-band magnitudes from the SDSS $gr$
magnitudes using the metal-poor Population II color transformation of
\citet{jor06}.\label{fig05}}
\end{figure*}

We plot in Figure~\ref{fig05} the apparent magnitude distribution of our
metal-poor candidate population.  There are 11,916 metal-poor candidates
in our sample with $V < 14$.  For comparison, at $V < 14$ there are 1,800
HES candidates in the list of \citet{chr08}, 1,559 candidates in the list
of \citet{fre06b}, and 125 HK candidates in the list of \citet{bee92}.
Given that $32.5^{+3.0}_{-2.9}$\% of the targets chosen with our
full selection have $-3.0 \lesssim \mathrm{[Fe/H]} \lesssim -2.0$
and $3.8^{+1.3}_{-1.1}$\% have [Fe/H] $\lesssim -3.0$, in our sample
one should be able to identify approximately 3,900 VMP stars and 450
EMP stars with $V < 14$.  APASS, 2MASS, and WISE data are available for
the entire sky, so it will be possible for the first time to search for
bright metal-poor stars in the entire northern sky as well as near the
southern Galactic plane.  Our candidate list is available upon request.

Of the giant stars targeted for moderate-resolution spectroscopy as
part of HES follow up, about $40.0^{+1.2}_{-1.2}$\% were found to have
$-3.0 \lesssim \mathrm{[Fe/H]} \lesssim -2.0$ and $3.9^{+0.5}_{-0.5}$\%
were found to have [Fe/H] $\lesssim -3$ \citep[e.g.,][]{sch09}.
Combining its yield with the apparent magnitude distribution of
its candidates, HES would be expected to find about 720 VMP giants
and 70 EMP giants with $V < 14$.  Of the apparently bright HES stars
observed with moderate-resolution spectroscopy by \citet{fre06b}, about
$8.7^{+0.7}_{-0.7}$\% were found to have  $-3.0 \lesssim \mathrm{[Fe/H]}
\lesssim -2.0$ and $1.5^{+0.3}_{-0.3}$\% were found to have [Fe/H]
$\lesssim -3$.  Even though our selection efficiency is currently slightly
less than HES, the larger number of candidates from our selection will
allow a survey based on our selection to identify a larger number of
apparently bright extremely metal-poor stars.

Our candidates are also bright in the sense that they are very luminous,
among the most luminous known metal-poor stars.  Using the scaling
relation

\begin{eqnarray}
L/L_{\odot} & = & (R/R_{\odot})^2 (T_{\mathrm{eff}}/T_{\mathrm{eff,\odot}})^4, \\
            & = & (M/M_{\odot}) (g/g_{\odot})^{-1} (T_{\mathrm{eff}}/T_{\mathrm{eff,\odot}})^4,
\end{eqnarray}

\noindent
and taking the mass of our stars as $0.8~M_{\odot}$, the bolometric
luminosity $L$ of our stars can be approximated as

\begin{eqnarray}
\log{\left(L/L_{\odot}\right)} & = & \log{0.8} - (\log{g} - 4.44) + 4\log{\left(T_{\mathrm{eff}}/5777~\mathrm{K}\right)}.\label{eq-lum}
\end{eqnarray}

\noindent
We find that 24 of our stars with $\mathrm{[Fe/H]} \lesssim -2.0$
have $\log{L/L_{\odot}} \gtrsim 3$.  That makes them more luminous
than all of the metal-poor stars analyzed by \citet{nor13a,nor13b} and
\citet{yon13a,yon13b} and comparable in brightness to the extremely
distant and luminous stars identified by \citet{boc14a,boc14b}
in the field and by several authors in dwarf spheroidal galaxies
\citep{kir08,fre10a,fre10b,sim10}.

\begin{figure*}
%\epsscale{0.8} % aastex
\plotone{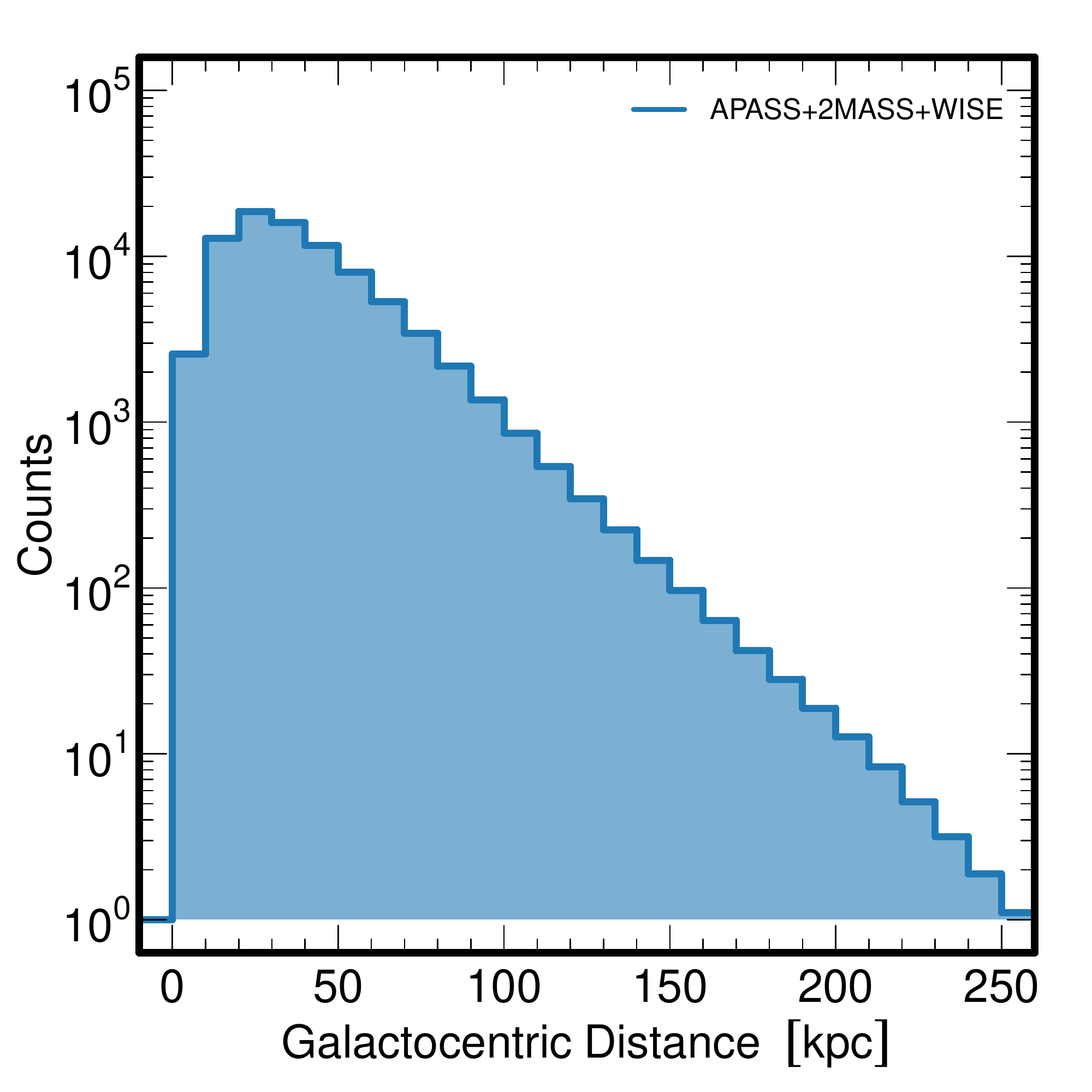}
\caption{Distance distribution of our candidate metal-poor stars.
The luminous nature of our candidates makes them excellent tracers of
the distant halo.\label{fig06}}
\end{figure*}

The bright absolute magnitudes of our metal-poor candidates suggests that
at fainter apparent magnitudes, our selection can provide a large number
of distant halo tracers.  Distant halo tracers are important for kinematic
estimates of the Milky Way's mass as well as the determination of the halo
metallicity profile \citep[e.g.,][]{bat05,bat06,xue08,xue14}. We determine
approximate distances for our Magellan/MIKE sample by interpolating a
[Fe/H] $= -2.5$, 10 Gyr Dartmouth isochrone \citep{dot08}.  We compute
absolute $W1$ band magnitudes as a function of bolometric luminosity $L$
from Equation (\ref{eq-lum}), then derive the distance modulus for each
of our metal-poor stars.  At 100 kpc, one of our luminous metal-poor
stars would have $V \approx 17.5$; near the Milky Way's virial radius
at 200 kpc, one of our luminous candidates would have $V \approx 19$.
To approximate the distance distribution of our entire sample, we convolve
the apparent magnitude distribution of our entire candidate sample with
the $W1$ absolute magnitude distribution of our Magellan/MIKE sample.
We plot the result in Figure~\ref{fig06}.  Cooler stars are also more
easily searched for signs of $r$-process enhancement than warmer stars,
so our metal-poor giants are a better sample to constrain the $r$-process
than a sample of warmer, metal-poor turnoff stars.  Likewise, since
cool stars typically have deeper lines at constant metallicity than warm
stars, cool stars are the only way to learn about the detailed chemical
compositions of the most metal-poor stars.

Should they persist, most of the Population III stars in the Galaxy
are thought to currently reside in the bulge \citep[e.g.,][]{tum10}.
Ground-based objective prism surveys are impractical in crowded bulge
fields, while near UV based photometric searches like SkyMapper are
effected by strong absolute and significant differential reddening.
On the other hand, our photometric selection operates well even in
crowded fields.  Similarly, since the effects of extinction and
reddening are $\sim\!\!50$ smaller in the WISE bands than in the
optical \citep[e.g.,][]{yua13}, our infrared selection is minimally
effected by absolute or differential reddening.  Indeed, if we take
the distance to the Galactic Center as $R_0 = 8.2$ kpc \citep{bov09}
and compute absolute $W1$ band magnitude as a function of $L$ for our
Magellan/MIKE sample, then two of our confirmed VMP stars are within
about 2 kpc of the Galactic Center: 2MASS J181503.64-375120.7 and
2MASS J183713.28-314109.3.  The former has [Fe/H] $= -2.85$ and the
latter has [Fe/H] $= -2.70$.  One of our confirmed EMP stars -- 2MASS
J155730.10-293922.7 -- is approximately 4 kpc from the Galactic Center.
These three stars are among the most metal-poor stars yet found in the
bulge \citep{nes13,gar13}.  Our selection can be extended to denser bulge
fields with existing, high-resolution VISTA/VIRCAM VVV near-infrared and
Spitzer/IRAC GLIMPSE mid-infrared data \citep{min10,sai12,ben03,chu09}.
In the future, grism spectroscopy with NIRCam on the James Webb Space
Telescope in bulge fields will be able to very efficiently select EMP
stars based on the lack of molecular absorption in $R \sim 2,\!000$
spectra between 2.4 and 5 microns \citep{rie05}.

\section{Conclusions}

We have developed a metal-poor star selection that uses only
public, all-sky APASS optical, 2MASS near-infrared, and WISE
mid-infrared photometry to identify bright metal-poor stars
through their lack of absorption near 4.6 microns.  Our selection
is efficient: our high-resolution follow-up observations confirmed
that $3.8^{+1.3}_{-1.1}$\% of our candidates have $\mathrm{[Fe/H]}
\lesssim -3.0$ and a further $32.5^{+3.0}_{-2.9}$\% have $-3.0 \lesssim
\mathrm{[Fe/H]} \lesssim -2.0$.  We used our selection to identify
seven previously unknown extremely metal-poor stars with $\mathrm{[Fe/H]}
\lesssim -3.0$ and a further 105 very metal-poor stars with $-3.0 \lesssim
\mathrm{[Fe/H]} \lesssim -2.0$.  In total, we identified 11,916 metal-poor
star candidates with $V < 14$, increasing by more than a factor of five
the number of publicly available metal-poor star candidates in this
range of apparent magnitude.  The bright apparent magnitudes of our
candidates will enable a large survey based on our candidate list to
significantly increase the number of extremely metal-poor stars with
detailed chemical abundance measurements.  Our infrared selection is
well suited to the identification of metal-poor stars in the bulge,
and two of our confirmed very metal-poor stars with $\mathrm{[Fe/H]}
\lesssim -2.7$ are within about 2 kpc of the Galactic Center.  They are
among the most metal-poor stars known in the bulge.

\acknowledgments
We thank Tim Beers, Anna Frebel, Jason Tumlinson, and Josh Winn.
We are especially grateful to Thomas Masseron for providing unpublished
carbon-enhanced synthetic spectra in the infrared.  This research has
made use of NASA's Astrophysics Data System Bibliographic Services and
both the SIMBAD database and VizieR catalog access tool, CDS, Strasbourg,
France.  The original description of the VizieR service was published
by \citet{och00}.  This research has made use of the NASA/IPAC Infrared
Science Archive, which is operated by the Jet Propulsion Laboratory,
California Institute of Technology, under contract with the National
Aeronautics and Space Administration.  This publication makes use of
data products from the Wide-field Infrared Survey Explorer, which is a
joint project of the University of California, Los Angeles, and the Jet
Propulsion Laboratory/California Institute of Technology, funded by the
National Aeronautics and Space Administration.  This publication makes
use of data products from the Two Micron All Sky Survey, which is a joint
project of the University of Massachusetts and the Infrared Processing and
Analysis Center/California Institute of Technology, funded by the National
Aeronautics and Space Administration and the National Science Foundation.
This research was made possible through the use of the AAVSO Photometric
All-Sky Survey (APASS), funded by the Robert Martin Ayers Sciences Fund.
This publication was partially based on observations obtained at the
Gemini Observatory, which is operated by the Association of Universities
for Research in Astronomy, Inc., under a cooperative agreement with the
NSF on behalf of the Gemini partnership: the National Science Foundation
(United States), the National Research Council (Canada), CONICYT
(Chile), the Australian Research Council (Australia), Minist\'{e}rio
da Ci\^{e}ncia, Tecnologia e Inova\c{c}\~{a}o (Brazil) and Ministerio
de Ciencia, Tecnolog\'{i}a e Innovaci\'{o}n Productiva (Argentina).
A.~R.~C acknowledges support through Australian Research Council Laureate
Fellowship LF0992131, from the Australian Prime Minister's Endeavour
Award fellowship for facilitating his research at MIT, and through
European Research Council grant 320360: The Gaia-ESO Milky Way Survey.
Support for this work was provided by the MIT Kavli Institute for
Astrophysics and Space Research through a Kavli Postdoctoral Fellowship.

{\it Facilities:} \facility{CTIO:2MASS}, \facility{FLWO:2MASS},
\facility{Gemini:South (GMOS-S spectrograph)}, \facility{Magellan:Clay
(MIKE spectrograph)}, \facility{Mayall (Echelle spectrograph)},
\facility{WISE}

\clearpage

\clearpage

\clearpage

\clearpage

\clearpage

\clearpage

\clearpage
\LongTables	% emulateapj
\begin{deluxetable}{lrrrrr}
\tablecaption{Metal-Poor Candidate Velocities and Stellar Parameters for Stars Observed with Gemini South/GMOS-S\label{tbl-1}}
\tablewidth{0pt}
\tablehead{\colhead{Object} & \colhead{$v_{los}$} & \colhead{$T_{\rm eff}$} & \colhead{$\log{g}$} & \colhead{[Fe/H]} & \colhead{[$\alpha$/Fe]} \\
\colhead{(2MASS)} & \colhead{(km s$^{-1}$)} & \colhead{(K)} & \colhead{} & \colhead{} & \colhead{}}
\startdata
J121325.53-380021.6 & $25.8^{5.2}_{-2.3}$       & $5339^{+64}_{-69}$    & $4.64^{+0.12}_{-0.11}$        & $-0.47^{+0.07}_{-0.07}$       & $0.24^{+0.04}_{-0.04}$ \\
J121812.33-372106.5 & $-26.3^{+27.8}_{-0.1}$    & $3499^{+41}_{-27}$    & $4.56^{+0.40}_{-0.16}$        & $0.77^{+0.18}_{-0.72}$        & $0.21^{+0.03}_{-0.01}$ \\
J122934.47-323307.4 & $231.2^{+0.2}_{-0.2}$     & $5002^{+9}_{-7}$              & $2.33^{+0.03}_{-0.03}$        & $-1.42^{+0.01}_{-0.01}$       & $0.44^{+0.02}_{-0.02}$ \\
J135216.59-355425.8 & $150.0^{+0.3}_{-0.3}$     & $5208^{+15}_{-17}$    & $3.01^{+0.04}_{-0.04}$        & $-1.62^{+0.02}_{-0.03}$       & $0.45^{+0.02}_{-0.03}$ \\
J135841.47-315110.2 & $5.6^{+0.3}_{-0.3}$       & $5256^{+17}_{-14}$    & $4.80^{+0.03}_{-0.03}$        & $-0.26^{+0.02}_{-0.02}$       & $0.15^{+0.01}_{-0.01}$ \\
J141733.51-274514.4 & $35.2^{+0.4}_{-0.4}$      & $4874^{+21}_{-20}$    & $3.81^{+0.05}_{-0.05}$        & $-0.24^{+0.02}_{-0.01}$       & $0.12^{+0.02}_{-0.02}$ \\
J141924.83-230737.1 & $39.3^{+0.3}_{-0.3}$      & $4766^{+24}_{-16}$    & $4.48^{+0.08}_{-0.05}$        & $-0.55^{+0.03}_{-0.03}$       & $0.06^{+0.01}_{-0.01}$ \\
J150159.91-261349.4 & $-107.3^{+1.2}_{-1.3}$    & $4969^{+44}_{-49}$    & $2.35^{+0.16}_{-0.13}$        & $-1.02^{+0.04}_{-0.07}$       & $0.41^{+0.05}_{-0.05}$ \\
\enddata
%\tablecomments{}
\end{deluxetable}

\clearpage
\LongTables	% emulateapj
\begin{deluxetable}{lrrrrrrcr}
%\rotate         % aastex
\tablecaption{Metal-Poor Candidate Velocities and Stellar Parameters for Stars Observed with Magellan/MIKE\label{tbl-2}}
\tablewidth{0pt}
\tablehead{\colhead{Object} & \colhead{$v_{los}$} & \colhead{$T_{\rm eff}$} & \colhead{$\log{g}$} & \colhead{[Fe/H]} & \colhead{$\xi$} & \colhead{[C/Fe]} & \colhead{Selection} & \colhead{Notes} \\
\colhead{(2MASS)} & \colhead{(km s$^{-1}$)} & \colhead{(K)} & \colhead{} & \colhead{} & \colhead{(km s$^{-1}$)} & \colhead{} & \colhead{} & \colhead{}}
\startdata
J122800.24-303137.6 & 50.4 & \dots & \dots & \dots & \dots & \dots & v1 & 1,2 \\
J122851.02-300335.0 & -91.8  & \dots & \dots & \dots & \dots & \dots & v1 & 3 \\
J122859.57-293619.2 & 7.7  & \dots & \dots & \dots & \dots & \dots & v1 & 1,2,4 \\
J122950.34-304017.2 & 85.1 & 4760  & 1.22  & -2.41 & 2.20 & \dots & v1 \\
J123015.17-305541.1 & 33.1 & \dots & \dots & \dots & \dots & \dots & v1 & 1,2 \\
J123100.36-300953.2 & 57.1 & \dots & \dots & \dots & \dots & \dots & v2 & 3 \\
J123118.86-311401.9 & 48.4 & \dots & \dots & \dots & \dots & \dots & v1 & 1 \\
J123122.75-301530.1 & 75.5 & 4860  & 1.74  & -2.21 & 1.50 & \dots  & v1 \\
\enddata
\tablecomments{1. Visual inspection revealed the target to be unambiguously metal-rich. 2. Ca II H \& K in emission. 3. Broad spectral features. 4. Evidence of binarity. }
\end{deluxetable}

\clearpage
\LongTables    % emulateapj
\begin{deluxetable}{lccccccccc}
%\rotate		% aastex
\tablecaption{Metal-Poor Candidate Star Positions and Photometry\label{tbl-3}}
\tablewidth{0pt}
\tablehead{\colhead{Object} & \colhead{RA} & \colhead{DEC} & \colhead{$V$} &
\colhead{$B-V$} & \colhead{$J$} & \colhead{$H$} & \colhead{$K$} & \colhead{$W1$} & \colhead{$W2$} \\
\colhead{(2MASS)} & \colhead{} & \colhead{} & \colhead{} &
\colhead{} & \colhead{} & \colhead{} & \colhead{} & \colhead{} & \colhead{}} 
\startdata
% cat table-photometry.txt
% cat gemini-stellar-parameters.txt | awk '{print $1, $9, $10, $5, $6, $7}'
% cat magellan-stellar-parameters+photometry-2014a.txt | awk '{print $17, $18, $19, $48, $49-$48, $35, $36, $37, $27, $28 }'
% cat magellan-stellar-parameters+photometry-2014b-halo.txt | awk '{print $17, $18, $19, $48, $49-$48, $35, $36, $37, $27, $28 }'
%#OBJECT & RA_1a & DEC_1a & johnson_vmag & b-v & j_m_2mass & h_m_2mass & k_m_2mass & w1mpro & w2mpro \\
J121325.53-380021.6     & 12:13:25.5    & -38:00:21     & 11.27 & 0.81  & 9.92  & 9.49  & 9.43  & 9.36  & 9.39 \\
J121812.33-372106.5     & 12:18:12.3    & -37:21:06     & 14.91 & 1.09  & 12.82 & 12.24 & 12.11 & 12.05 & 12.06 \\
J122800.24-303137.6     & 12:28:00.2    & -30:31:38     & 14.12 & 1.12  & 11.94 & 11.35 & 11.24 & 11.17 & 11.20 \\
J122851.02-300335.0     & 12:28:51.0    & -30:03:35     & 13.17 & 0.92  & 11.24 & 10.75 & 10.60 & 10.61 & 10.60 \\
J122859.57-293619.2     & 12:28:59.6    & -29:36:19     & 13.29 & 1.29  & 10.64 & 10.07 & 9.84  & 9.79  & 9.77 \\
J122934.47-323307.4     & 12:29:34.4    & -32:33:07     & 12.59 & 0.97  & 10.62 & 10.09 & 9.97  & 9.91  & 9.94 \\
J122950.34-304017.2     & 12:29:50.3    & -30:40:17     & 12.84 & 0.84  & 11.07 & 10.57 & 10.46 & 10.40 & 10.42 \\
J123015.17-305541.1     & 12:30:15.2    & -30:55:41     & 13.07 & 0.87  & 11.46 & 10.98 & 10.92 & 10.87 & 10.89 \\
\enddata
%\tablecomments{}
\end{deluxetable}

\clearpage
\begin{deluxetable}{llcccr}
\tablecaption{Previously Known Metal-Poor Stars in Our Follow-Up Target List\label{tbl-4}}
\tablewidth{0pt}
\tablehead{\colhead{Object} & \colhead{Alias} &\colhead{$T_{\rm eff}$} &
\colhead{$\log{g}$} & \colhead{[Fe/H]} & \colhead{Reference} \\
\colhead{(2MASS)} & \colhead{} &\colhead{(K)} & \colhead{} & \colhead{} & \colhead{}}
\startdata
J121846.49-161050.8 & HE 1216-1554 & \dots & \dots & $-3.33$ & \citet{fre06b} \\
J123916.23-264746.8 & ~~~~~~~~~~~\dots & \dots & \dots & $-2.23$ & \citet{har96} \\
J123918.93-264941.9 & ~~~~~~~~~~~\dots & \dots & \dots & $-2.23$ & \citet{har96} \\
J125045.87-282656.3 & ~~~~~~~~~~~\dots & \dots & \dots & $-2.01$ & \citet{mor90} \\
J131947.00-042310.2 & HE 1317-0407 &  4525 &  0.30 & $-3.10$ & \citet{hol11} \\
J143355.92-124035.7 & HE 1431-1227 & \dots & \dots & $-2.58$ & \citet{fre06b} \\
J212138.43-395612.9 & BPS CS 30492-0116 & \dots & \dots & $-2.05$ & \citet{bee00} \\
J220216.36-053648.5 & HE 2159-0551 & \dots & \dots & $-3.03$ & \citet{fre06b} \\
J220841.10-394512.2 & BPS CS 22881-0040 & \dots & \dots & $-2.27$ & \citet{bee00} \\
J221059.73-022525.2 & HE 2208-0240 & \dots & \dots & $-2.06$ & \citet{fre06b} \\
J222025.81-102320.2 & BPS CS 22886-0042 & \dots & \dots & $-2.61$ & \citet{aok07} \\
J230449.13-424348.0 & HE 2302-4259 & \dots & \dots & $-3.15$ & \citet{fre06b} \\
J231300.04-450706.6 & HE 2310-4523 & \dots & \dots & $-2.60$ & \citet{fre06b} \\
J232607.41-055006.9 & BPS CS 22949-0048 & 4620 & 0.95 & $-3.37$ & \citet{roe14}
\enddata
\end{deluxetable}

\appendix
\section{Data Quality Flags}

We make four WISE data quality checks.  First, that the WISE $W1$, $W2$,
and $W3$ photometry be free of artifacts.  Second, that the objects are
fully consistent with a point source.  Third, that the quality of the
photometry in both $W1$ and $W2$ has been rated `A'.  Fourth, that the
level of contamination by the Moon in $W1$, $W2$, and $W3$ be consistent
with zero.  The following SQL commands can be used to reproduce our
initial selection by setting limits on an ``All Sky Search" of the
AllWISE Source catalog available from the NASA/IPAC Infrared Science
Archive\footnote{\url{http://irsa.ipac.caltech.edu/Missions/wise.html}}.

\begin{verbatim}
j_m_2mass - h_m_2mass between 0.45 and 0.6
and w3mpro > 8
and w1mpro - w2mpro between -0.04 and 0.04
and j_m_2mass - w2mpro >= 0.5
and cc_flags like '000_'
and ext_flg = 0
and ph_qual like 'AA__'
and moon_lev like '000_'
\end{verbatim}

\end{document}